\documentclass[page-classic]{epl2}
\usepackage{amsmath}
\usepackage{amssymb}
\usepackage{amsfonts}

\title{Might EPR particles communicate through a wormhole?}
\shorttitle{EPR correlations via a wormhole} %Insert here a short version of the title if it exceeds 70 characters

\author{E. Sergio Santini\inst{1,2}\thanks{santini@cbpf.br}}
\shortauthor{E.S. Santini}

\institute{ 
\inst{1} Instituto de Cosmologia, Relatividade e Astrof\'{\i}sica ICRA-BR  \\
Centro Brasileiro de Pesquisas F\'{\i}sicas \\
Rua Dr. Xavier Sigaud 150, Urca 22290-180 -- Rio de Janeiro, RJ -- Brasil  \\
\inst{2}  
Comiss\~ao Nacional de Energia Nuclear \\ 
Rua General Severiano 90, Botafogo 22290-901 -- Rio de Janeiro, RJ -- Brasil                
}

\abstract{
We consider the two-particle wave function of an Einstein-Podolsky-Rosen system, given by a two dimensional relativistic scalar field model. The Bohm-de Broglie interpretation is applied and the quantum potential is viewed as modifying the Minkowski geometry. In this  way an effective metric, which is analogous to a black hole metric in some limited region, is obtained in one case and a particular metric with singularities  appears in the other case, opening the possibility, following Holland,  of  interpreting the EPR correlations as being originated by an effective wormhole  geometry, through which the physical signals can propagate.}

\begin{document}

\maketitle

\section{Introduction}
There is an increasing interest in the application of the Bohm - de Broglie (BdB) interpretation of quantum mechanics to  several areas, such as quantum  cosmology, quantum gravity and quantum field theory, see for example \cite{must}\cite{cons}\cite{fshojai}\cite{alves}\cite{nikolic}. In this work, we develop a causal approach to the Einstein-Podolsky-Rosen (EPR) problem i.e. a two-particle correlated system. We attack the problem from the point of view of quantum field theory, considering the two-particle function for a scalar field.  In the BdB approach, it is possible to interpret the quantum effects as modifying the geometry in such a way that the scalar particles see an effective geometry. 
As a first example, we show that a two dimensional EPR model, in a particular quantum state and under a non-tachyonic approximating condition, can exhibit an effective metric that is analogous  to a two dimensional black hole (BH) in some region (which is limited by the approximations we made).
In a second example, for a two-dimensional static EPR model we are able to show that quantum effects produce an effective geometry with  singularities in the metric, a key ingredient of a bridge construction or wormhole. 
In this way, and following a suggestion by Holland \cite{holland}, we can envisage the possibility of interpreting the EPR correlations as driven by an effective wormhole, through which physical signals can propagate.
 This letter is organized as follows: in the next section we recall the basic features of a relativistic scalar field and  write the two-particle wave equation. Then, we apply the BdB interpretation to it and, from the generalized Hamilton-Jacobi equation, we visualize the quantum potential as generating an effective metric. Having done that, we next study two dimensional EPR models and show how the effective metric appears, being a BH metric (in some region) in the first example and a particular metric with singularities in the second. The last section is for the conclusions.

\section{ Scalar field theory and its BdB interpretation}\label{2}

The Schr\"{o}dinger  functional  equation for a quantum relativistic 
scalar field is given by    
\begin{equation}
\label{qsf}
i \hbar \frac{\partial \Psi (\phi ,t)}{\partial t} = 
\int d^3x \biggr\{\frac{1}{2}\biggr[-\hbar ^2  
\frac{\delta ^2}{\delta \phi ^2} +
(\nabla \phi)^2\biggl] + U(\phi) \biggl\} \Psi (\phi ,t) ,
\end{equation}
where $\Psi (\phi ,t)$ is a functional with respect to $\phi(\mathbf{x})$ and a function  with respect to $t$. A normalized solution $\Psi (\phi ,t)$ can be expanded as

\begin{equation}
\Psi[\phi,t]=\sum_{n=0}^{\infty} \int d^{3}k_{1}...d^{3}k_{n}c_{n}(\vec{\mathbf{k}}^{n},t)\Psi_{n,\vec{\mathbf{k}}^{n}}[\phi]
\end{equation}
where $\vec{\mathbf{k}}^{(n)}\equiv\{\mathbf{k}_{1},...\mathbf{k}_{n}\}$ and $\mathbf{k}_{j}$ is the momenta of particle $j$, being the functionals $\Psi_{n,\vec{\mathbf{k}}^{(n)}}[\phi]$  a complete orthonormal basis.

For free fields, the $n$-particle wave function  is given by (see for example \cite{schweber}\cite{long})

\begin{equation}\label{np}
\psi_{n}(\vec{\mathbf{x}}^{(n)},t)=\langle 0|\hat{\phi}(t,\mathbf{x_1})...\hat{\phi}(t,\mathbf{x_n})|\Psi\rangle,
\end{equation}
where $\vec{\mathbf{x}}^{(n)}\equiv \{\mathbf{x}_{1},...\mathbf{x}_{n}\}$.

The wave function (\ref{np}) satisifies

\begin{equation}
\sum_{j=0}^{n}[(\partial^{\mu}\partial_{\mu})_{j}+\frac{m^2 c^2}{\hbar^2}]\psi_{n}(\vec{\mathbf{x}}^{(n)},t) = 0.
\end{equation}

For the two-particle wave function we have

\begin{equation}
\sum_{j=1}^{2}[(\partial^{\mu}\partial_{\mu})_{j}+\frac{m^2 c^2}{\hbar^2}]\psi_{2}(\vec{\mathbf{x}}^{(2)},t) = 0
\end{equation}
which is

%\begin{widetext}
\begin{equation}
\label{2p}
[(\partial^{\mu}\partial_{\mu})_{1}+\frac{m^2 c^2}{\hbar^2}]\psi_{2}(\mathbf{x_1},\mathbf{x_2},t) + [(\partial^{\mu}\partial_{\mu})_{2}+\frac{m^2 c^2}{\hbar^2}]\psi_{2}(\mathbf{x_1},\mathbf{x_2},t)= 0 .
\end{equation}
%\end{widetext}
%\begin{floatequation}
%\mbox{\textit{see eq. ~\eqref{2p}}}
%\end{floatequation}

In order to apply the BdB interpretation, we substitute $\psi_2=R \exp(iS/\hbar)$ in eq.(\ref{2p}) obtaining two equations, one for the real part and the other for the imaginary part. The first equation reads

\begin{equation}\label{2pr}
\partial^{\mu_1}S\partial_{\mu_1}S - m^2 c^2 -\hbar^2\frac{ (\partial^{\mu}\partial_{\mu})_{1} R }{R} + 
\partial^{\mu_2}S\partial_{\mu_2}S - m^2 c^2 -\hbar^2\frac{ (\partial^{\mu}\partial_{\mu})_{2} R }{R}
= 0
\end{equation}
that can be written as

\begin{equation}\label{2pr2}
\eta^{\mu_{1}\nu_{1}}\partial_{\mu_1}S\partial_{\nu_1}S
+\eta^{\mu_{2}\nu_{2}}\partial_{\mu_2}S\partial_{\nu_2}S = 2 c^2 {\cal M}^2
 \end{equation}
where $\eta^{\mu \nu}$ is the Minkowski metric and

\begin{equation}\label{M2}
{\cal M}^2\equiv m^2 (1-\frac{Q}{2 m^2 c^2})
\end{equation}
with

\begin{equation}\label{qpot}
Q \equiv -\hbar^2\frac{ (\partial^{\mu}\partial_{\mu})_{1} R }{R}-\hbar^2\frac{ (\partial^{\mu}\partial_{\mu})_{2} R }{R} .
\end{equation}

The equation that comes from the imaginary part is

\begin{equation}\label{continu}
\eta^{\mu_{1}\nu_{1}}\partial_{\mu_1}(R^2\partial_{\nu_1}S)+\eta^{\mu_{2}\nu_{2}}\partial_{\mu_2}(R^2\partial_{\nu_2}S)=0
\end{equation}
which is a continuity equation.

Equation (\ref{2pr2}) is the Hamilton-Jacobi equation for a 2-particle system of mass $2{\cal M}$. The term $Q$ is the quantum potential whose effect can be interpreted as a modification of the system's mass with respect to its classical value $2 m$. We see that ${\cal M}^2$ is not positive-definite, a feature that is associated whith the existence of tachyonic solutions. To overcome this problem one can, for example,  choose  initial conditions in such a way that a positive ${\cal M}^2$ value is obtained for an initial time. Because of the continuity equation, this will be true all the time.

Now, following an idea proposed by  L. De Broglie \cite{debroglie} and fruitfully applied to gravity in \cite{fshojai} and \cite{alves}, we can rewrite the Hamilton-Jacobi equation (\ref{2pr2}) as

\begin{equation}\label{2pre}
\frac{\eta^{\mu_{1}\nu_{1}}}{(1-\frac{Q}{2 m^2 c^2})}\partial_{\mu_1}S\partial_{\nu_1}S
+\frac{\eta^{\mu_{2}\nu_{2}}}{(1-\frac{Q}{2 m^2 c^2})}\partial_{\mu_2}S\partial_{\nu_2}S =
2 m^2 c^2 .
\end{equation}

We can interpret the quantum effects as realizing a conformal transformation of the Minkowski metric $\eta^{\mu\nu}$ in such a way that the effective metric is given by

\begin{equation}\label{curved}
g_{\mu\nu}= (1-\frac{Q}{2 m^2 c^2})\eta_{\mu\nu}
\end{equation}
and eq. (\ref{2pre}) can be written as

\begin{equation}\label{2prc}
{\cal D}_{\mu_{1}}S{\cal D}^{\mu_{1}}S
+{\cal D}_{\mu_{2}}S{\cal D}^{\mu_{2}}S = 2 m^2 c^{2}
 \end{equation}
where $D_{\mu}$ stands for a covariant differentiation with respect to  the metric $g_{\mu\nu}$ and  $\partial_{\mu}S={\cal D}_{\mu}S$, because $S$ is a scalar.

Then, as was already shown by Shojai et al. in \cite{fshojai}, the quantum potential  modifies the background geometry giving a curved space-time with the metric given by eq. (\ref{curved}). In some appearances, according to Shojai, space-time geometry shows a dual aspect: it sometimes looks like (semiclassical) gravity and sometimes looks like quantum effects.

\section{Two dimensional models}

Two dimensional models have been studied for a long time, in order to address subjects such  as gravitational collapse, black holes and quantum effects. We analyse two dimensional models because some aspects in a low dimensional model have the same  behavior as the more realistic four-dimensional models \footnote{Furthermore, low dimensional models appear naturally  in effective string theories.}.
In this section, we are going to show two examples in two dimensions of a two-particle EPR system that exhibit an effective metric, as in eq.(\ref{curved}).  Because of the singularities of this effective metric, it resembles a two dimensional BH type solution, as  presented in \cite{alves} and \cite{mann}, and this is the key that could allow us to connect the EPR correlations with an effective wormhole geometry. Before  presenting the examples, we briefly recall the basic features of that solution.

The two dimensional BH presented in \cite{mann} consists of a point particle situated at the origin, with density $\rho=\frac{M}{2\pi G_N}\delta(x)$, where $M$ is the mass of the particle and $G_N$ is the Newton gravitational constant. A symmetric solution of the field equation of this problem, without cosmological constant (see\cite{mann} section 3), is given by the metric:

\begin{equation}\label{metric}
ds^2=-(2M|x|-C)dt^2+\frac{dx^2}{2M|x|-C}
\end{equation}
where $C$ is a constant.
The sign of the quantity $\alpha \equiv 2M|x|-C$ determines the type of region: timelike regions are for positive $\alpha$ and spacelike regions are for negative $\alpha$. The points at which  $\alpha(x)=0$ are coordinate singularities and locate the event horizons of the space, which in the present case are at:

\begin{equation}
|x|=\frac{C}{2M} .
\end{equation}
The horizons only exist if $C$ and $M$ are of the same sign. For example, for positive $C$ and positive $M$, there are two horizons, at  $x_h=\mp\frac{C}{2M}$, with the source located in a spacelike region surrounded by two timelike regions. In particular, if $C=0$ and $M>0$ there is only one horizon at $x=0$ surrounded by a timelike region. The metric (\ref{metric}) can be cast in ``conformal coordinates'' $(t,y)$ (see\cite{mann} section 3). For example, in the case of $M$ and $C$ positives, the transformation $x=\frac{C+e^{2My}}{2M}$ , for $x \in (\frac{C}{2M}, \infty)$, transform the metric in:

\begin{equation}
ds^2=e^{2My}(-dt^2+dy^2) .
\end{equation}

In the following part of this section we consider  two dimensional  models for the EPR problem.

First example: non-tachyonic EPR model. For the two-particle system, we obtained a conformal transformation of the metric where the conformal factor is associated with the quantum potential, eq.(\ref{curved}). We shall deal here with the non-tachyonic case, i.e. we need to impose the possitivity of ${\cal M}^2$. One way to do this is to assume that $Q$ is a small perturbation, viewing  eq.(\ref{M2}) as an approximation, to first order in $\frac{Q}{2 m^2 c^2}$, of an exponential: 

\begin{equation}
{\cal M}^2=m^2 \exp(-\frac{Q}{2 m^2 c^2})\simeq m^2 (1-\frac{Q}{2 m^2 c^2})
\end{equation}
which is valid for 
\begin{equation}\label{approx}
|\frac{Q}{2 m^2 c^2}|\ll 1 .
\end{equation}

Then, with this assumption, which means a very restricted example, we have for the effective metric, eq.(\ref{curved}):

\begin{equation}\label{curvedexp}
g_{\mu\nu}= \exp(-\frac{Q}{2 m^2 c^2})\eta_{\mu\nu} .
\end{equation}

Now we assume that our two dimensional two-particle system satisfies an EPR 
condition, i.e. their positions $x_1$ and $x_2$ are correlated in such a way that $x_1+x_2=constant$\cite{einstein}. Then, the dependence of the amplitude $R$ (and of the quantum potential  $Q$) on the coordinates  $x_1$ and $x_2$, can be cast as a function of only one coordinate, say $x_1$, and defining $z\equiv x_1$ and assuming that the amplitude of the state is independent of time (see below, eq.(\ref{Airy})), we can write, with a little abuse of notation, $R=R(x_1,x_2)=R(z)$ and $Q=Q(x_1,x_2)=Q(z)$.

The line element now becomes 

\begin{equation}\label{elementexp}
ds^2=\exp{(-\frac{Q(z)}{2 m^2 c^2})}(-dt^2+dz^2) .
\end{equation}

We are now going to make the assumption that the quantum entangled state is prepared in such a way that its amplitude is given by the (real) Airy function\cite{jeffreys}:

\begin{equation}\label{Airy}
R(z)=\frac{A}{\pi}\int_{0}^{\infty}\cos\left (-s\left (z+\frac{K \hbar^2}{2M m^2c^2}\right )\left (\frac{2M m^2c^2}{\hbar^2}\right )^{1/3}+\frac{s^3}{3}                  \right )ds
\end{equation}
where $K$ and $M$ are  integration constants (positive, for instance) and $A$ is a normalization constant. Then the amplitude $R(z)$ satisfies the equation 

\begin{equation}\label{airyequation}
\frac{d^2R}{dz^2}+(K+\frac{2M m^2c^2}{\hbar^2}z)R=0
\end{equation}
and taking into account  the definition of the quantum potential, eq.(\ref{qpot}), we can see that $Q$ satisfies

\begin{equation}\label{linearQ}
\frac{dQ}{dz}=4 M m^2 c^2  .
\end{equation}

Defining a coordinate transformation from $z$ to $y$ by mean $2My=-\frac{Q(z)}{2 m^2 c^2}$ 
the line element (\ref{elementexp}) in $(t,y)$ "conformal coordinates" reads

\begin{equation}
ds^2= e^{2My}(-dt^2+dy^2) .
\end{equation}

Now we can make a coordinate transformation from $y$ to $x$ by mean

\begin{equation}
2Mx=C + e^{2My}
\end{equation}
where $C$ is a constant and $x \in (\frac{C}{2M}, \infty)$. The line element $ds^2$ in terms of the $(t,x)$ coordinates  is given now by

\begin{equation}\label{bhsimilar}
ds^2=-(2M|x|-C)dt^2+\frac{dx^2}{2M|x|-C}
\end{equation}
where we made a symmetrical extension for the other values of $x$ other than $(\frac{C}{2M}, \infty)$ (see \cite{mann}).
Hence, we arrive at the same metric defining a two dimensional BH type solution, eq.(\ref{metric}). In spite of this similarity, we must stress that the metric for the analysed EPR problem is given by eq.(\ref{bhsimilar})  only when the approximation given by 
eq.(\ref{approx}) and the assumption (\ref{Airy}) are satisfied, which means only for $x$ in the region defined  by   
\begin{equation}\label{region}
2M|x|-C = e^{-2Mz}, \, \, \, with \, \, \, |2Mz|<<1
\end{equation}
or 
\begin{equation}\label{region2}
 x \in (\frac{1+C}{2M}-\epsilon,\frac{1+C}{2M}+\epsilon)
\end{equation}
being $\epsilon$ a  constant  satisfying  $0<\epsilon <<\frac{1}{2M}$.
(Here we used the particular form of $Q$,  $Q=4 M m^2 c^2z$, that comes from (\ref{airyequation}) with $K=0$). In fact, the extension we made, led us beyond the region of validity of our approximation, for $2M|x|-C$ very different from $1$. Hence we can consider this particular EPR problem as an analog model of a BH, only in some limited region which is given by (\ref{region2}). 
 The coordinate singularities are located at $x=\mp\frac{C}{2M}$, at the poles of the quantum potential $Q(z)$, but they are outside the region (\ref{region2}) and  the approximation (\ref{approx}) breaks down. 
This makes our present example very limited. Let us now consider our next example. 

Second example: a static model. In the last part of this section we shall show a very simple example  where  singularities are present in the transformed metric. We consider again the two-particle wave function of a scalar field in two dimensions. Following the approach of Alves in \cite{alves} we shall  see that, for the static  case, it is possible to obtain a solution as a metric of the curved space-time (the effective metric), which comes from Eqs. (\ref{2pr2}) and (\ref{continu}). 
In the present case, these equations are:

\begin{equation}\label{2prtwo}
\eta^{11}\partial_{x_1}S\partial_{x_1}S + 
\eta^{11}\partial_{x_2}S\partial_{x_2}S = 2 m^2 c^2 (1-\frac{Q}{2 m^2 c^2})
\end{equation}

\begin{equation}\label{cont-two}
\partial_{x_1}(R^2\partial_{x_1}S)+\partial_{x_2}(R^2\partial_{x_2}S)=0
\end{equation}

Now we consider that our two-particle system satisfies the EPR condition $p_1=-p_2$ which in the BdB interpretation, using the Bohm guidance equation $ p=\partial_{x}S$, can be written as

\begin{equation}
\partial_{x_1}S=-\partial_{x_2}S .
\end{equation}

Using this condition in eq. (\ref{cont-two}), we have

\begin{equation}\label{contBdB}
\partial_{x_1}(R^2\partial_{x_1}S)=\partial_{x_2}(R^2\partial_{x_1}S)
\end{equation}
and this equation has the solution

\begin{equation}\label{G}
R^2\frac{\partial S}{\partial x_1}= G(x_1+x_2)
\end{equation}
where $G$ is an arbitrary (well behaved) function of $x_1+x_2$.

Substituting eq.(\ref{G}) in eq.(\ref{2prtwo}), we have

\begin{equation}\label{qpot2}
 2 m^2 c^2 (1-\frac{Q}{2 m^2 c^2})= 2 \left (\frac{G}{R^2}\right )^2
\end{equation}
and taking into account the expression (\ref{qpot}) for the quantum potential, the last equation reads

\begin{equation}\label{nonlinear}
8G^2 + \hbar^2 (\partial_{x_1}(R^2))^2-\hbar^2 2 R^2\partial_{x_1}^{2}R^2 +\hbar^2(\partial_{x_2}(R^2))^2-\hbar^2 2 R^2\partial_{x_2}^{2}R^2 -8m^2 c^2 R^4 = 0 .
\end{equation}
A solution of this nonlinear equation is 
\begin{equation}
 R^2 = \frac{1}{2m^2 c^2}\left (C_{1} \sin\left (\frac{m c}{\hbar}(x_1+x_2)\right )+C_2\right )
\end{equation}
provided a suitable  function $G(x_1+x_2)$, which can be obtained from (\ref{nonlinear}) by substituting the solution.

In order to interpret the effect of the quantum potential, we can re-write eq.  (\ref{2prtwo}) using (\ref{qpot2}) obtaining

\begin{equation}
\eta^{11}\partial_{x_1}S\partial_{x_1}S + 
\eta^{11}\partial_{x_2}S\partial_{x_2}S = 2 \left (\frac{G}{R^2}\right )^2
\end{equation}
or

\begin{equation}
m^2\frac{\eta^{11}}{(\frac{G}{R^2})^2}\partial_{x_1}S\partial_{x_1}S + 
m^2\frac{\eta^{11}}{(\frac{G}{R^2})^2}\partial_{x_2}S\partial_{x_2}S = 2m^2
\end{equation}
that we write as

\begin{equation}
g^{11}\partial_{x_1}S\partial_{x_1}S + 
g^{11}\partial_{x_2}S\partial_{x_2}S = 2m^2 c^2
\end{equation}
and then  we see that the quantum potential was "absorbed" in the new metric $g_{11}$, which is:

\begin{eqnarray}\label{m}
&g_{11}=\frac{1}{g^{11}}=\frac{\eta_{11}}{c^2 m^2}(\frac{G}{R^2})^2= \nonumber \\ &\frac{1}{4}\frac{2C_{1}^2 \sin^2\left (\frac{m c}{\hbar}(x_1+x_2)\right )-C_{1}^2\cos^2\left (\frac{mc}{\hbar}(x_1+x_2)\right )-2C_{1}C_{2}\sin\left (\frac{m c}{\hbar}(x_1+x_2)\right )}{\left (C_{1}\sin\left (\frac{mc}{\hbar}(x_1+x_2)\right )+C_{2}\right )^2} .&
\end{eqnarray}

We can see that this  metric is singular at the zeroes of the denominator in (\ref{m}). According to the model reviewed at the begining of this section, this is characteristic of a two dimensional BH solution (see \cite{alves} and \cite{mann} ). Then our two-particle system  "sees" an effective  metric with singularities, a fundamental component of a  wormhole \cite{visser}, through which the physical signals can propagate \footnote{It is interesting to note that a wormhole coming from a  (Euclidean) conformally flat metric with singularities was shown by  Hawking \cite{hawwormholes}. Consider the metric: 
\begin{equation}
ds^2=\Omega^2dx^2
\end{equation}
with
\begin{equation}
\Omega^2=1+\frac{b^2}{(x-x_{0})^2} .
\end{equation}
This looks like a metric with a singularity at $x_{0}$. However, the divergence of the conformal factor can be thought as  the space opening out to another  asymptotically flat region connected with the first through a  wormhole of size $2b$.}.

\section{Conclusion}
We studied the  two-particle state of a scalar field under the EPR condition for the two dimensional case, in two situations, a non-tachyonic case and a static one. We found that the quantum potential can be interpreted as realizing a conformal transformation of the Minkowski metric to an  effective metric. In  the first situation, this effective metric is analogous to a BH metric in some limited region  and in the second situation the metric contains singularities, a key ingredient  of a bridge construction or wormhole. This opens the possibility, following a suggestion by Holland \cite{holland}, of interpreting the EPR correlations of the entangled particles as driven by an effective wormhole. Obviously, a more realistic (i.e. four dimensional) and more  sophisticated model (i.e. including the spin of the particles) must be studied. 
\acknowledgments
I would like to thank   Prof. Nelson Pinto-Neto from ICRA/CBPF, Prof. Sebasti\~ao Alves Dias from LAFEX/CBPF, Prof. Marcelo Alves from IF/UFRJ,  and the 'Pequeno Seminario' of ICRA/CBPF for  their useful comments.
I would also like to thank the Minist\'erio da Ci\^encia e Tecnologia/ CNEN and CBPF of Brazil for their financial support. Special thanks to the anonymous referee for the valuable criticism and corrections.

\end{document}